\documentclass[aps,prb,twocolumn,showpacs]{revtex4}

\usepackage{graphicx}

\newcommand{\edgetype}[1]{\ensuremath{\langle#1\rangle}}
\newcommand{\pedges}{\edgetype{311}}
\newcommand{\psides}{\{211\}}

\begin{document}

\title{Image deformation in field ion microscopy of faceted crystals}

\author{Daniel Niewieczerza{\l}}

\author{Czes{\l}aw Oleksy}

\affiliation{Institute of Theoretical Physics, University of Wroc{\l}aw,
Plac Maksa Borna 9, 50-204 Wroc{\l}aw, Poland}

\author{Andrzej Szczepkowicz}

\affiliation{Institute of Experimental Physics, University of Wroc{\l}aw,
Plac Maksa Borna 9, 50-204 Wroc{\l}aw, Poland}

\date{\today}

\begin{abstract}
We perform detailed numerical simulations of field ion microscopy images 
of faceted crystals and compare them with experimental observations. 
In contrast to the case of crystals with a smooth surface, 
for a faceted topography we find extreme deformations of the ion image.
Local magnification is highly inhomogeneous and may vary by an order of magnitude: from 0.64 to 6.7.
Moreover, the anisotropy of the magnification at a point located 
on the facet edge may reach a factor of 10.
\end{abstract}

\pacs%
{%
 68.37.Vj, 
 02.70.Dh, 
 02.70.Ns 
}

\maketitle

\section{Introduction\label{section-introduction}}

In recent years field ion microscopy (FIM) \cite{FIM-Tsong,FIM-Oxford} has found a number of applications
connected with thermally faceted crystals \cite{Madey1999,Madey1999a}. The most prominent example is the
research on the fabrication of atomically sharp electrodes 
\cite{FuChengNienTsong2001,Lucier2005,Rezeq2006,BrylSzczepkowicz2006,Kuo2006,FujitaShimoyama2008,Rahman2008}
and their use as practical electron \cite{Chang2009} and ion 
\cite{Kuo2008} point sources, including
laser-driven femtosecond electron sources \cite{Hommelhoff2006,Hommelhoff2009}. 
In these studies FIM is often
used to determine the electrode shape and the degree of its sharpness.
Another example of the application of FIM to faceted crystals is its use in
surface science studies of the faceting process itself, in its steplike 
and global form \cite{SzczepkowiczCiszewski2002,SzczepkowiczBryl2004,Szczepkowiczetal2005}, 
and in the study of the equilibrium crystal shape \cite{SzczepkowiczBryl2005}.
Worth mentioning are also FIM studies on single atom diffusion \cite{AntczakEhrlich2005} 
-- although not carried out on thermally-faceted surfaces, they also encounter the problem of large
image distortion at the facet edge.

It is well established that image magnification in FIM is not uniform \cite{FIM-Oxford}. This constitutes
a problem in the interpretation of the FIM images, and in the past researchers have tried to 
devise methods which would allow to take into account these effects. Both analytical 
and numerical calculations of the electric field and ion trajectories have been carried out 
\cite{FIM-Oxford,Vurpillot2001,Vurpillot2004}.
However, none of these calculations apply to faceted crystals, which are recently most
often examined in FIM. Our work is aimed at filling this gap. Faceted crystals, with their 
atomically sharp edges, produce very large local variations of the intensity and direction
of the electric field. This produces huge deflections in ion trajectories,
and, consequently, huge distortions in the resulting microscopic image.

In this paper we perform detailed numerical calculations of the ion trajectories 
in the FIM of faceted crystals. This leads to surprising insights into the 
interpretation of field ion images of such crystals. 
Where possible, we confront our numerical results with available experimental data 
concerning image distortion.

\section{Method of calculation\label{Calculation}}

In this section we discuss the numerical approach to simulation of
field ion microscope images of faceted surfaces. First, we find
the electric field distribution in a model of FIM by solving
Laplace's equation, and then the ion trajectories are obtained
form classical equations of motion.

\subsection{Construction of the model of a faceted crystal}

We will  use  specimens  generated earlier in Monte Carlo
simulation of adsorbate-induced faceting on curved surfaces of bcc
metal \cite{dn_czo}. A typical specimen has spherical shape with
faceted area around the the $[111]$ pole limited by the
inclination angle $\Theta =20^\circ$. The faceted region of a
specimen has a form of a single pyramid with \{112\} faces or it
contains step-like \{112\} mircofacets. We  also study truncated
pyramids and a pyramid with double edges.

\begin{figure}
\centering
\includegraphics[width=3cm]{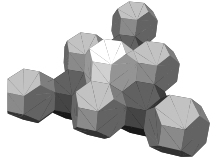}
\caption{ 3D view of the apex of a pyramid  formed of the
truncated octahedrons. } \label{figure1}
\end{figure}

We are especially interested in answering the question how the
morphology of faceted surface affects the electric field
distribution near the emitter surface. Moreover we want to examine
the local magnification on images of  faceted surfaces. Thus the
FIM model should take into account the atomic roughness of the tip
surface and assure long enough distance between the emitter and
the screen, or in other words, the model should span many length
scales. Such conditions cause big difficulties in numerical
solution of  Laplace equation on a discrete mesh. On the one hand,
the size of the mesh should be equal to the distance between the
tip and the screen ( $\sim$10~cm in real experiment). On the other
hand, the mesh should have resolution smaller than {1~\AA}  near
the tip. In this paper we demonstrate that solution of Laplace's
equation, in a case when  the distance between the emitter and the
screen is million times larger than the mesh resolution, is
possible assuming the following approximations:
\begin{itemize}
    \item The spherical emitter  with faceted region.
Atoms in the faceted region are represented by the truncated
octahedrons -- the Wigner-Seitz cells of the bcc crystal structure
(see Fig.~\ref{figure1}). The remaining part of the emitter is
approximated by  a smooth sphere of radius $R_0$.
    \item The spherical screen of radius $R_1$ which is about three hundred times
grater than $R_0$.
\end{itemize}
Let us comment the the choice of truncated octahedrons to
represent atoms in the  FIM model. We considered 3 possibilities:
(i) point like atoms and the surface represented by triangles,
(ii) hard sphere model, (iii) each atom  represented by the
truncated octahedron - the Wigner-Seitz cell of a bcc crystal
structure. All these cases were preliminarily tested in
simulations and we have chosen the  representation by truncated
octahedrons. The simplest representation by surface triangulation
does not accurately reflect the atomic roughness of (111), (211)
and (110) faces. The hard sphere model seems to be the best one,
but it is difficult to use.  If one takes the radius from the
close-packed  bcc structure, the surface contains plenty of holes
between the spheres. On the other hand, choice of  larger sphere
radius leads to the reduction of the atomic roughness.
Representation of atoms by truncated octahedrons ensures that the
bcc crystal faces have proper symmetry and atomic roughness, and
it is easy to use  in numerical calculation.

\subsection{Calculation of the electric field}

To  calculate the electric field distribution in the FIM model,
first one has to solve Laplace's equation (LE) in the space
between the emitter and the screen
\cite{Vurpillot2001,Vurpillot2004}.
\begin{equation}\label{LE}
\triangle V(\mathbf{r}) =0,
\end{equation}
where $V(\mathbf{r})$ is the electric potential at the point
\textbf{r}. It is assumed that $V=V_0$ at all points of the emiter
surface and the $V=V_1$ at all points of the screen. Then the
electric  field is  obtained:
\begin{equation}\label{gradE}
\mathbf{E}(\mathbf{r})=-\nabla V({\mathbf{r}}).
\end{equation}

To solve LE,  Eq.~(\ref{LE}), in the space $\Omega$ between the
emitter and the  screen, finite element method (FEM) is applied,
because it is well-suited  for simulation of geometrically
complicated domains \cite{zienkiewicz,Rahman2006}. All FEM related
calculations are performed using Getfem++ package \cite{getfem},
whereas tetrahedral meshes are  generated using TetGen library
\cite{tetgen}. In our application of the finite element method the
domain $\Omega$ is divided into a number of tetrahedrons and
Lagrange-type interpolation functions are used as the basic
functions.  For such mesh the electric field is constant in a
tetrahedron. Hence, to assure appropriate accuracy of \textbf{E},
especially near the emitter surface, the linear size of a
tetrahedron should be much smaller than the lattice constant $a$.
Direct computation does not allow for such mesh refinement, so we
overcome this difficulty in the following way. First, we combine
the numerical solution of LE with the analytical one. It follows
from our preliminary calculation that at distances  $r$ from  the
emitter greater than $5R_0$ the electrostatic potential $V(r)$ for
the faceted emitter is practically the same as for the spherical
emitter.

It is easy to show that the electrostatic potential $V_S$   for
the spherical emitter is given by the formula
\begin{equation}\label{V_spherical}
V_S(r)=\frac{V_1 R_1- V_0 R_0}{R_1-R_0}+
\frac{(V_0-V_1)R_0R_1}{R_1-R_0}\frac{1}{r}
\end{equation}
and applying  Eq.~(\ref{gradE}) gives the electric field
$\mathbf{E}_S$
\begin{equation}\label{E_spherical}
\mathbf{E}_S(\mathbf{r})=\frac{(V_0-V_1)R_0R_1}{R_1-R_0}\frac{\mathbf{r}}{r^3}
\end{equation}
A numerical solution of  LE  is limited to a subdomain
$\Omega^\prime$ of spherical shape with a radius
$R_{\mathrm{num}}$.
We apply Dirichlet boundary conditions to surfaces in
$\Omega^\prime$ with $V=V_0$ on internal (emitter) surface and
$V=V_S(R_{\mathrm{num}})$ on the external surface. In the space
outside the $\Omega^\prime$, i.e., for distance $r>
R_{\mathrm{num}}$, the analytical solution given by
Eq.~(\ref{V_spherical}) and Eq.~(\ref{E_spherical}) will be used.

To obtain a numerical solution of LE in the subdomain
$\Omega^\prime$ with appropriate accuracy  for FIM images
simulation, we use 3-step procedure described below:
\begin{enumerate}
\item  The domain $\Omega^\prime$ is divided into $n$ spherical
layers -- subdomains $\Omega_i$, $i=1, 2, \ldots, n$, each with
its own mesh of tetrahedrons. Then the LE is solved in the whole
$\Omega^\prime$ to get  values of  V on boundaries of each
subdomain. \item In each  subdomain $\Omega_i$  the LE is solved
again on a finer mesh of tetrahedrons applying Dirichlet boundary
conditions to $\Omega_i$ surfaces where the electric potential is
known from solution obtained in the previous step. \item To
calculate the electric field at a point \textbf{r} within
$\Omega_i$:
\begin{enumerate}
\item A local subdomain $\omega_\textbf{r}$ is constructed by
collecting all tetrahedrons with distance from \textbf{r} smaller
than $R_{\mathrm{loc}}$. \item A new  mesh is created by filling
with newly generated tetrahedrons the space inside the
$\omega_\textbf{r}$ boundary. There is  a special, regular
tetrahedron $T_\textbf{r}$ with  center located at $\textbf{r}$
\item Using this mesh, LE is solved imposing V values from the
$\omega_\textbf{r}$ boundary as Dirichlet condition. \item
Finally, the electric field is obtained as minus gradient of V
calculated  on   $T_\textbf{r}$
\end{enumerate}
\end{enumerate}

Typical technical data used in our calculation: the radius of
emitter $R_0=268$~\AA, assuming the lattice constant for tungsten:
a=3.16~\AA, the radius of $\Omega^\prime$ domain
$R_{\mathrm{num}}=33R_0$, the radius of the screen $R_1=267R_0$,
the number of subdomains $n=9$, the number of tetrahedrons used in
whole $\Omega^\prime$ for the 1st step is $N_T=4728000$ and for
the 2nd step  in each subdomain $\Omega_i$, $N_T>2.6*10^6$. The
size of the special tetrahedron $T_\textbf{r}$ is $\frac{a}{70}$
in the vicinity of emitter surface. Hence, the ratio of $R_1$ to
size of the special tetrahedron is over $10^6$.

It is worth to emphasize an important role of the third step in
our procedure of calculating the electric field by using a local
subdomain $\omega_\textbf{r}$. Omitting this step leads to an
incorrect  solution even in $\Omega_1$ subdomain,  although
$\Omega_1$ has the smallest linear size ($\sim1.07R_0$) and large
number of tetrahedrons in its mesh $N_T=3.4\times 10^6$.

We performed  several tests to verify presented method of solving
LE. The most important one concerned calculation of the electric
field for an ideal spherical emitter because the analytical
solution is known in this case -- Eq.~(\ref{E_spherical}). The
obtained numerical results are in good agreement with  analytical
solution. In the  examined distance range up to $r=8R_0$, the
relative error $|E(r)-E_S(r)|/E_S(r)$ is smaller than $5*10^{-4}$
and the deviation of $\mathbf{E}$ direction  from the radial
direction is smaller then $0.1^\circ$. We also obtained positive
results for checking the symmetry of the electric field
distribution on the different emitter shapes: pyramidal tip and
bcc spherical one.

\subsection{Calculation of the ion trajectories}

Motion  of an ion of mass {\em M} and charge  {\em q} under the
influence of force $\mathbf{F}(\mathbf{r})=q
\mathbf{E}(\mathbf{r})$ is described by the equation of motion
\begin{equation}\label{Newton}
M\frac{d^2\mathbf{r}}{dt^2}=q \mathbf{E}(\mathbf{r}).
\end{equation}
It is  assumed that the ion is formed with zero initial velocity
at a small distance ($a\frac{\sqrt3}{2}$) from a surface atom. It
is convenient to express quantities occurring  in
Eq.~(\ref{Newton}) in reduced units. Choosing lattice constant $a$
as unit of length, $(V_0-V_1)/a$ as unit of E and
$t_0=\sqrt{\frac{Ma^2}{q(V_0-V_1)}}$  as unit of time, the
equation  of motion Eq.~(\ref{Newton}) can be expressed  in the
form
\begin{equation}\label{Newt_red}
\frac{d^2\mathbf{\tilde{r}}}{d\tilde{t}^2}=\mathbf{\tilde{E}}(\mathbf{\tilde{r}}),
\end{equation}
where tilde denotes quantity in reduced unit. Thus, the
trajectories in reduced units do not depend on the applied
voltage, mass and charge of the ion. These quantities affect unit
of time $t_0$. To solve  Eq.~(\ref{Newt_red}) we apply the
velocity Verlet algorithm -- commonly used in molecular dynamics
simulations\cite{frenkel_MD}, with a time step $0.1t_0$.

\begin{figure}
\centering
\includegraphics[width=6cm]{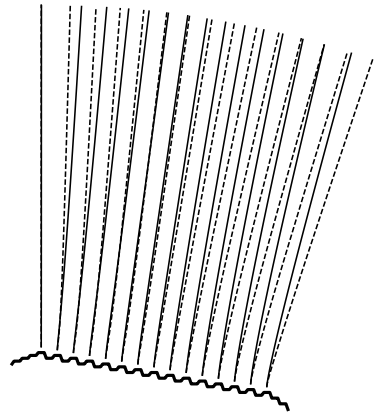}
\caption{ An example of ion trajectories (continuous lines) and
electric field lines (dashed lines) starting from the edge of a
pyramid. In this image the length of these lines is limited to $0.5R_0$.} 
\label{traj}
\end{figure}

It is obvious that the direction of ion trajectory is not
coincident with the electric field lines (see Fig.\ \ref{traj})
even in the case of the initial ion velocity equal to zero. Due to
the curvature of the electric field lines,  instantaneous ion
velocity at $\mathbf{r}$ is deflected from
$\mathbf{E}(\mathbf{r})$ up to large distances from the tip
surface.

\subsection{Calculation of a local magnification}

Knowing the full ion trajectories we can investigate magnification
not only on the distant microscope screen but also on virtual
screens at small distances from the emitter. Analysis of FIM
images on virtual screens is used to determine the minimal emitter
-- screen distance.
 Magnification $f$ is defined here as
\begin{equation}
f=\frac{ d^\prime}{d},
\end{equation}
 where $d$ is the distance of a pair of points on the specimen and
$d^\prime$ is the corresponding distance on the screen.

Another important quantity is the local magnification $f_l$ of
distance $d$ defined as
\begin{equation}
f_l=\frac{\frac{ d^\prime}{d}}{f_{av}},
\end{equation}
where $f_{\mathrm{av}}$ stands for average magnification. The
average magnification can be replaced by magnification of a large
distance, eg.  the distance between ends of pyramid edges.

We found that local magnification of faceted crystal has a
long-range dependence on the distance $r$ from the tip. Using the
tip with curvature radius $R_0 \approx 27$~nm  we obtained that
the local magnification reaches $73\%$ of its final  value $f_l^s$
at $r=R_0$.  At larger distances $f_l(5R_0)\approx0.9f_l^s$,
$f_l(10R_0)\approx0.94f_l^s$, and $f_l(30R_0)\approx0.98f_l^s$.
Hence, to control the error of  $f_l$, the screen --tip distance
in the numerical model should be at least 30 times greater than the tip
curvature radius, which is much greater than the distances used in
previous numerical studies (see
eg.\cite{Vurpillot2001,Vurpillot2004}).

\section{Results and discussion\label{Results}}

\subsection{A perfect, single atom pyramid\label{perf}}

\begin{figure}
	\includegraphics{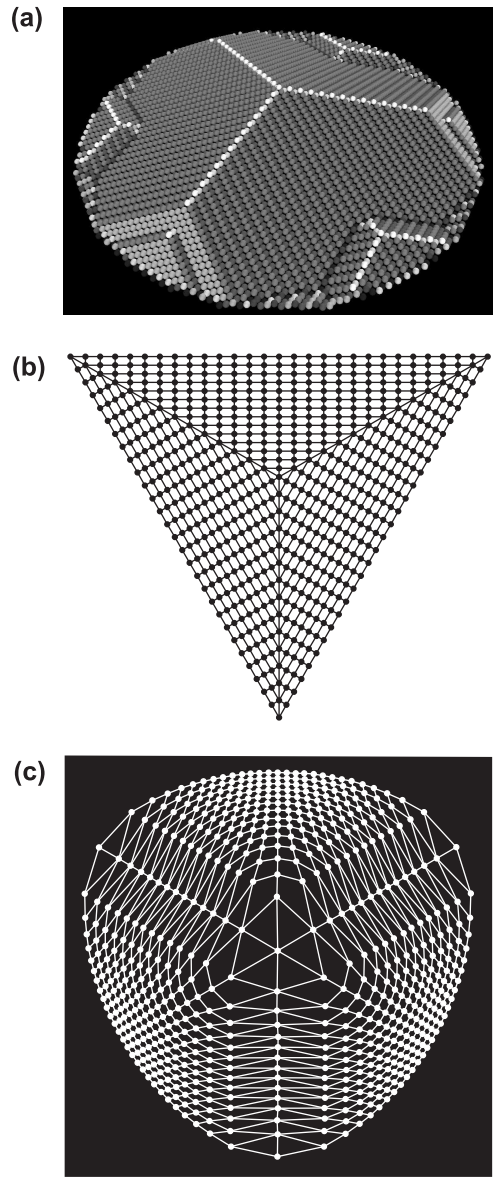}
	\caption{
	A single atom pyramid.
	(a)~A three-dimensional visualization of the sample.
	(b)~A projection of the surface atoms of the sample onto a plane.
	(c)~Calculated projection onto the microscope screen following the charged-particle trajectories in the electric field.
	}
	\label{piramida-1-6-15-ostra}
\end{figure}

We begin our calculation with the simple case of a perfect 3-sided pyramid pointing in the [111] direction,
formed by three densely packed crystal facets: (211), (121), (112). The crystallographic direction of the pyramid edges is \pedges. We assume that all atoms are in the BCC lattice positions and that the pyramid is ended by a single atom. This
implies that that the number of atoms in the succcesive (111) planes is $1,6,15,28\ldots$.
The assumed sample configuration is shown in Fig.~\ref{piramida-1-6-15-ostra}(a) and (b). 
After calculating the resulting electric field $\vec E(\vec r)$ between the sample and the microscope screen,
we map the surface atoms onto the microscope screen by following the classical charged-particle trajectories (ions or electrons) -- the solution of Eq.~\ref{Newton}. The result is shown in Fig.~\ref{piramida-1-6-15-ostra}(c). 
The calculated image shows some striking features.
\begin{enumerate}

\item\label{A-straight} Of all the straight lines seen on the sample [Fig.~\ref{piramida-1-6-15-ostra}(b)], only the pyramid
edges remain straight lines on the image [Fig.~\ref{piramida-1-6-15-ostra}(c)].

\item\label{A-apex} The magnification of the image is highly inhomogenous. There is a great enhancement of the magnification
at the apex of the pyramid. The local magnification factor at the apex is 4.5.

\item\label{A-edge-parallel} Along the pyramid edge, the local parallel magnification varies from 2.8 near the apex, reaching a minimum (0.64) near the center of the edge (de-magnification), increasing to a value of 1.5 near the pyramid base.

\item\label{A-edge-transverse} Along the pyramid edge, the local transverse magnification is 7.4 near the apex, quickly decreasing to a constant value of 6.7. 

\item\label{A-edge-anisotropy} As seen from the comparison of features \ref{A-edge-parallel}.\ and \ref{A-edge-transverse}., the magnification of the image is exceedingly anisotropic -- the image is extremely stretched in one direction. In the middle of the pyramid edge, the anisotropy is $6.7/0.64\sim 10$.

\end{enumerate}

\begin{figure}
	\includegraphics{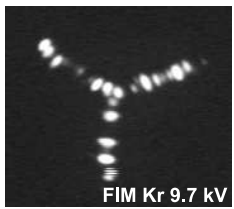}	
	\caption{A FIM image of oxygen-covered tungsten pyramid \cite{SzczepkowiczBryl2005}.}
	\label{FIM-Kr-ostra}
\end{figure}

It would be very interesting to confront the above results with experiment. However, there is no easy way to calibrate
an FIM image in terms of real distances. It is possible with careful FIM of samples with low average radii, where individual atoms are resolved, but as the average radius increases, the microscope resolution decreases and one looses the 
natural reference length of the lattice constant. However, observation \ref{A-edge-anisotropy}. appears to explain 
a peculiar feature of FIM images of pyramids -- the images of atoms at the edges are often not circular, but stretched in the direction perpendicular to the edges -- see Fig.~\ref{FIM-Kr-ostra}.

\begin{figure}
	\includegraphics{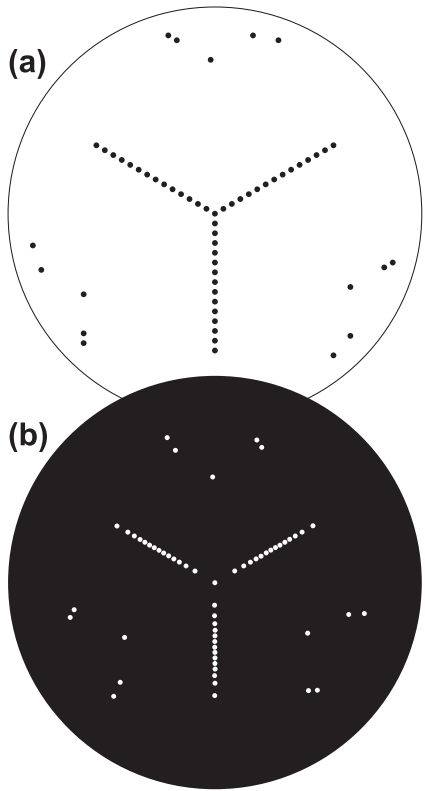}	
	\caption{Atoms which satisfy the ionization criterion (above a treshold field): 
	(a) the sample, (b) the simulated FIM image.}
	\label{piramida-p}
\end{figure}

The image of atoms shown in Fig.~\ref{piramida-1-6-15-ostra}(c) does not correspond directly to the experimentally observed
FIM image, because in the real microscope only the most protruding atoms are visible. One of the reasons is that the protruding atoms, which can be thought of as ``sharp'' features on the otherwise smooth surface, cause local enhancement in the electric field. Imaging requires field ionization, which takes place for electric field magnitudes exceeding the ionization treshold. Experiments show that for the crystal shape considered here, only the edges of the pyramid are seen in the FIM image. For this reason, for the realistic simulation of the image, we select the atoms
where the electric field above the atom exceeds a certain treshold [Fig.~\ref{piramida-p}(a)]. 
In the last step we calculated the ion trajectories starting from the seletected atoms and extending up to the microscope screen to generate a simulated FIM image [Fig.~\ref{piramida-p}(b)].

\subsection{Truncated pyramids\label{trunc}}

\begin{figure}
	\includegraphics{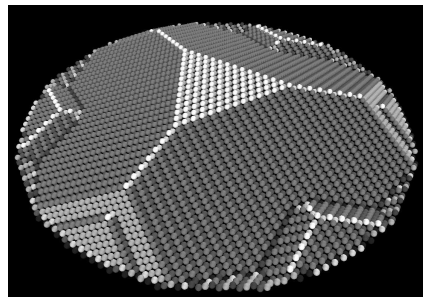}	
	\caption{Visualization of a truncated pyramid (7 atomic (111) layers removed).}
	\label{piramida-1-6-15-scieta-PC7}
\end{figure}

\begin{figure*}
	\includegraphics{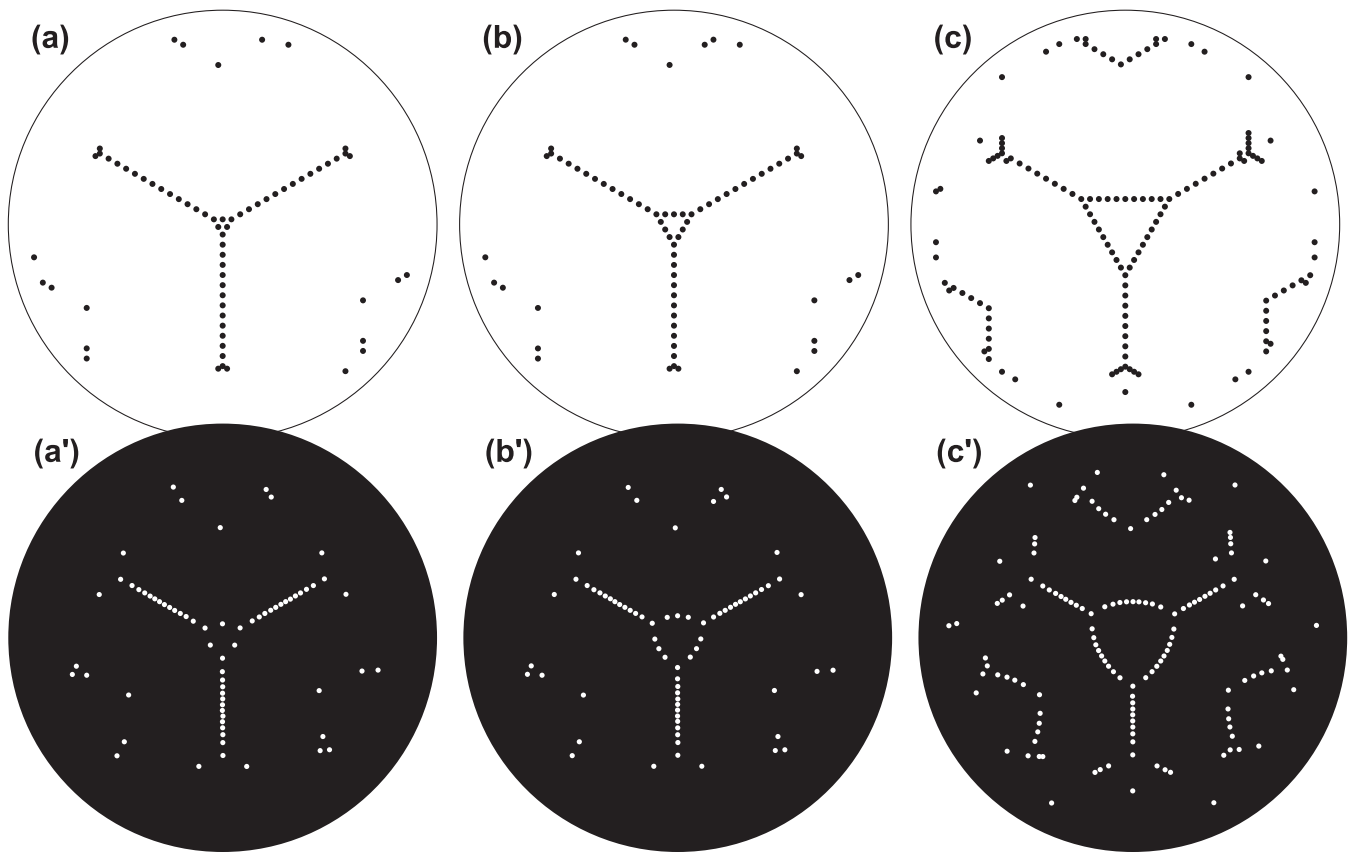}	
	\caption{Truncated pyramids with (a) 1, (b) 2, and (c) 5 atomic layers removed.
	The corresponding simulated ion images are shown in $(a')$--$(c')$}
	\label{piramidy-PC-1-2-5}
\end{figure*}

\begin{figure}
	\includegraphics{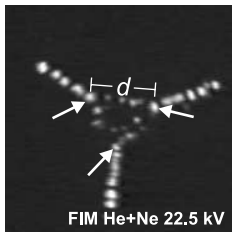}	
	\caption{A FIM image of a truncated pyramid. The arrows point to the edge end points.}
	\label{FIM-HeNe-tepa}
\end{figure}

The exact shape of the \psides-faceted crystal can be controlled by careful temperature
treatment. It is possible to obtain pyramids with various
degrees of vertex truncation or vertex rounding, which can be quantified by measuring
the distance $d$ between the pyramid edge end points\cite{SzczepkowiczBryl2005}. However, such
measurement in FIM suffers from errors caused by the inhomogenity of the local magnification.
If one assumes constant magnification in FIM, the measurements of $d$ are always overestimated
-- that is, the pyramids are actually more ``sharp'' than they appear to be in the microscopic image.
To calculate the correction for the measurements carried out in \cite{SzczepkowiczBryl2005},
we calculated a series of images of truncated pyramids. The height of the single atom 
pyramid was 14 geometrical  atomic layers; now we remove 1,2,3 etc, obtaining an atomic
configuration shown in Fig.~\ref{piramida-1-6-15-scieta-PC7}. 
Three examples of resulting images are shown in Fig. \ref{piramidy-PC-1-2-5}.
A striking feature is the shape deformation of the truncated vertex area, which has
a triangular boundary on the assumed sample, but almost a circular boundary on the simulated image. 
In FIM experiments,
the truncated area is not well resolved, but the boundary of the truncated vertex indeed often
appears curved outword, as shown in Fig.~\ref{FIM-HeNe-tepa}.
As could be expected from the results presented in Sect.~\ref{perf}, the local magnification
in the central part of the image is high. For one atomic layer removed, the area of the truncated vertex increases
8.0 times; for 2 layers -- 4.8 times; for 5 layers -- 2.1 times.
The corresponding linear magnifications, proportional to the square roots, are 
2.8, 2.2 and 1.4, respectively. As the total height of the pyramid was 14 layers,
the local magnifications calculated above correspond to height truncation of 7\%, 14\% and 36\%.

\begin{figure}
	\includegraphics{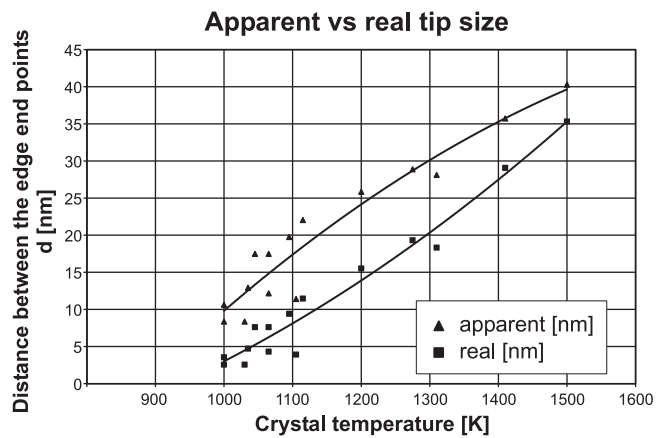}	
	\caption{The corrected dependence of the equilibrium crystal shape on the temperature for oxygen-covered tungsten \cite{SzczepkowiczBryl2005}.}
	\label{wykres-T-d}
\end{figure}

Applying similar considerations as above to the magnification
of the distance $d$ between the edge end points, it is possible to correct previous work
on the dependence of the equilibrium crystal shape on the temperature \cite{SzczepkowiczBryl2005},
as shown in Fig.~\ref{wykres-T-d}.
Note that without the correction based on this work, the amount of truncation of the pyramid is
highly overestimated.

\begin{figure}
	\includegraphics{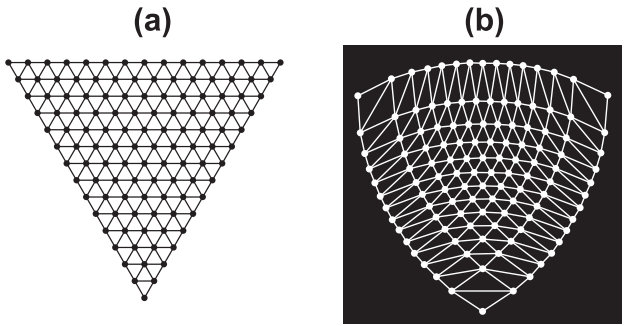}	
	\caption{(a)~The atomic arrangement on the (111) plane of the truncated pyramid in Fig.~\ref{piramida-1-6-15-scieta-PC7}.
	(c)~Calculated projection onto the microscope screen following the charged-particle trajectories in the electric field.}
	\label{piramida-1-6-15-scieta-PC7-obraz-scianki-111}
\end{figure}

In Fig.~\ref{piramida-1-6-15-scieta-PC7-obraz-scianki-111} we demonstrate the simulated image of the (111) surface
formed by truncating half of the height of the \psides\ pyramid. 
In the experiment, at this pyramid size
the indywidual atoms would not be resolved; however, in principle all the (111) lattice sites could be visualized
using single atom adsorption on the (111) plane, similarly as in the work of Antczak and Ehrlich \cite{AntczakEhrlich2005}.
In this way they have observed huge deformations of the image of an atomically flat crystal facet: curving of the
atomic lines and an increase of the magnification near the boundary of the facet (Fig. 2 in Ref. \cite{AntczakEhrlich2005}). 
Both of these features are present in the result of our simulation 
shown in Fig.~\ref{piramida-1-6-15-scieta-PC7-obraz-scianki-111}, further confirming the validity of our model.

\subsection{A steplike-faceted crystal (hill-and-valley faceting)}

\begin{figure}
	\includegraphics{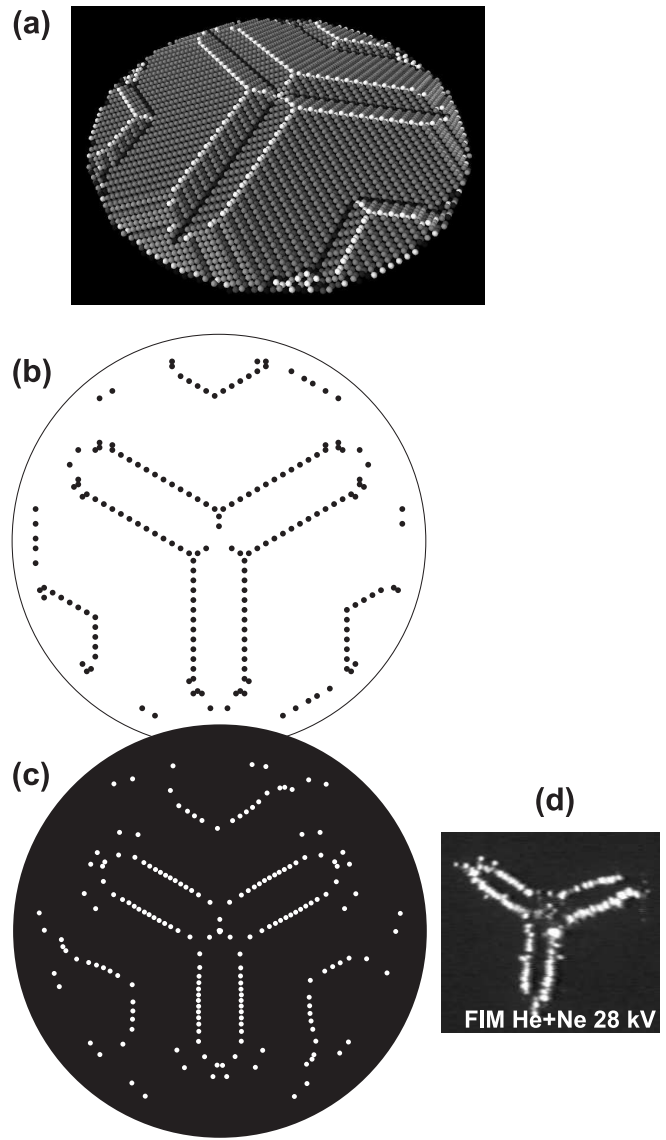}	
  \caption{A steplike-faceted crystal. (a)~3D visualization of the assumed atomic configuration.
	(b),(c) The positions of the image atoms on the sample and on the screen. (d) Comparison with a real image.}
	\label{piramida-e}
\end{figure}

In the experiment, occurence of global faceting, where the shape of the faceted crystal is convex,
such as described in Sect. \ref{perf} and \ref{trunc}, is only possible under special conditions:
small crystal, high annealing temperature, and high desorption temperature of the adsorbate.
If these conditions are not fulfilled, one obtains a steplike-faceted crystal
\cite{Madey1999,Szczepkowiczetal2005}. This is due to the kinetic limitations on the 
surface diffusion of the crystal material. For this reason we consider here a steplike-faceted
sample, as shown in Fig.~\ref{piramida-e}. 
The assumed atomic configuration closely corresponds
to surface topography observed in the experiments (see e.g. Ref.\cite{SzczepkowiczBryl2004}).

Comparison of the simulated and real FIM image [Fig.~\ref{piramida-e}(c) and (d)] shows an
overall agreement, but also a discrepancy in one aspect. In the experiment, the six \pedges\ step edges are not pairwise
parallel. 
One possible reason is that our model sample has a different topography than the real sample 
near the boundary of the faceted region.
Another possibility is that the real atomic configuration is not as perfect
as assumed in the calculation [Fig.~\ref{piramida-e}(a)]. Note the high variation of the local magnification
along the step edge in the simulated image. Unfortunately, this effect cannot be verified
in the real image [Fig.~\ref{piramida-e}(d)] due to the lack of single-atom resolution.

\begin{figure}
	\includegraphics{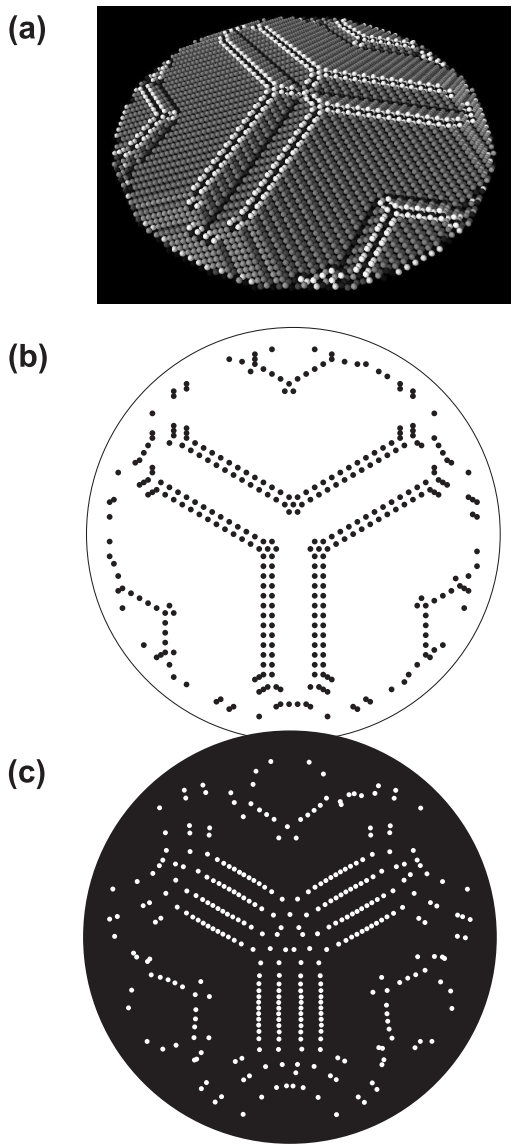}	
	\caption{(a),(b) The assumed atomic configuration and (c) the simulated ion image, corresponding to Pd/W faceting.}
	\label{piramida-ed}
\end{figure}

\begin{figure}
\includegraphics{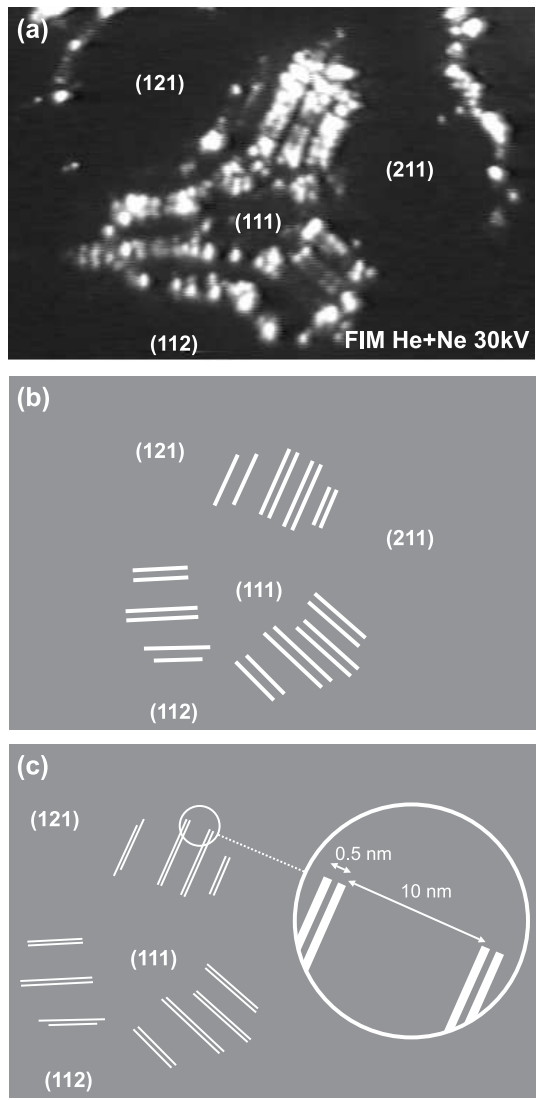}%
\caption%
{%
 (a)~FIM image of a faceted palladium-covered tungsten surface \cite{Szczepkowiczetal2005}.
 (b)~The position \pedges\ edges in image (a).
 (c)~Approximate actual position of the edges on the crystal surface. The distances are estimated
 from FIM and STM experiments \cite{Szczepkowiczetal2005}.
 \label{palladium-fim}%
}
\end{figure}

Another atomic configuration, which is interesting for comparison with experiment, is shown in Fig.~\ref{piramida-ed}.
This is significant for the case of palladium-induced faceting of tungsten \cite{NienMadey1997,Szczepkowiczetal2005}.
In these experiments, palladium forms a pseudomorphic physical monolayer on the sample. The
\pedges\ edges of pyramids/steps are truncated, with one atomic row removed, as shown in Fig.~\ref{piramida-ed}(a).
Comparison of the actual atomic arrangement [Fig.\ref{piramida-ed}(b)] 
with the simulated FIM image at the microscope screen [Fig.\ref{piramida-ed}(c)] reveals a serious problem
in the interpretation of the FIM image in such a case: it is very difficult to infer from 
the image on the screen [Fig.\ref{piramida-ed}(c)]
the actual configuration of the atoms [Fig.\ref{piramida-ed}(b)].
The image observed on the screen is very misleading, suggesting the presence of three series 
of four almost equally-spaced parallel edges. This constituted a serious problem in the interpretation of the ion images
of Pd/W faceting. Our present calculations finally fully confirm the interpretation of the ion images assumed 
in previous work\cite{SzczepkowiczCiszewski2002,Szczepkowiczetal2005}, illustrated here in Fig.~\ref{palladium-fim}.
Calculations yield a very large magnification anisotropy: in the middle of the facet edges,
the local transverse magnification is 3.1, while the local parallel magnification is 0.74.
This results in the image magnification anisotropy of 4.2 at this point -- the FIM image
is stretched over 4 times in the direction transverse to the facet edges. In the experiment,
the local transverse magnification is probably even larger, a rough estimate yields a factor of  
7--8 (compare Fig.~\ref{palladium-fim}). However, this number is only an estimate, as there
is no direct way to measure local magnification in the FIM experiment.

\section{Conclusion}

We performed detailed trajectory calculations in the FIM of \psides-faceted crystals 
(the case often encountered in research), 
and compared them with available experimental data. We found that
the experimentally obtained FIM images of faceted crystals are often very misleading, 
and one has to rely on numerical calculations to fully account for 
huge local deformations of the microscopic image. Our main conclusions can be summarised as follows:

\begin{enumerate}

\item In the consideration of faceted crystals, it is necessary for the model of FIM  
to take into account the atomic roughness of the tip surface and, at the same time, 
the large distance from 
the tip to the screen -- the model must span many length scales.
Our solution of Laplace's equation demonstrates that the distance between the tip 
and the screen in the FIM model should be at least 30 times greater than the 
tip curvature radius in order to calculate the local magnification correctly.

\item A combination of the finite element method with an analytical solution allows for 
accurate calculation of the electric field in the space between the tip and the screen.

\item The FIM images of faceted crystals exhibit very large variations of the local magnification.
We have found local magnification factors as high as 6.7 and as low as 0.64 (local de-magnification).

\item As one moves along the pyramid or step edge, the parallel magnification varies
by a factor of 2--4, reaching a minimum near the middle of an edge.

\item\label{anisotropyx} The images of pyramid edges or step edges are exceedingly stretched in 
the direction perpendicular to the edge. The anisotropy of the local magnification,
defined as the ratio of local magnification in two ortogonal directions, reaches
a factor of 10 (for a perfect \psides\ pyramid).

\item As a consequence of \ref{anisotropyx}., an image of a truncated facet edge 
mimics the image of two distant parallel edges. Great caution is required
in the interpretation of this kind of FIM images.

\item The image of a single flat facet is distorted in such a way that 
the local magnification increases as one moves toward the facet edge.
The atomic rows do not appear straight on the image. Although we have calculated
only the case of a (111) facet on top of a \psides-pyramid, comparison with
experiment suggests that this conclusion is generally true for isolated flat facets.

\item Truncated \psides\ pyramids appear significantly more truncated in the FIM image 
than they are in reality; this effect is most pronounced for small truncations.
Numerical calculations are necessary to estimate the size of a
\psides-faceted tip.

\end{enumerate}


\begin{thebibliography}{29}
\expandafter\ifx\csname natexlab\endcsname\relax\def\natexlab#1{#1}\fi
\expandafter\ifx\csname bibnamefont\endcsname\relax
  \def\bibnamefont#1{#1}\fi
\expandafter\ifx\csname bibfnamefont\endcsname\relax
  \def\bibfnamefont#1{#1}\fi
\expandafter\ifx\csname citenamefont\endcsname\relax
  \def\citenamefont#1{#1}\fi
\expandafter\ifx\csname url\endcsname\relax
  \def\url#1{\texttt{#1}}\fi
\expandafter\ifx\csname urlprefix\endcsname\relax\def\urlprefix{URL }\fi
\providecommand{\bibinfo}[2]{#2}
\providecommand{\eprint}[2][]{\url{#2}}

\bibitem[{\citenamefont{Tsong}(1990)}]{FIM-Tsong}
\bibinfo{author}{\bibfnamefont{T.~T.} \bibnamefont{Tsong}},
  \emph{\bibinfo{title}{Atom-probe field ion microscopy}}
  (\bibinfo{publisher}{Cambridge University Press}, \bibinfo{year}{1990}).

\bibitem[{\citenamefont{Miller et~al.}(1996)\citenamefont{Miller, Cerezo,
  Hetherington, and Smith}}]{FIM-Oxford}
\bibinfo{author}{\bibfnamefont{M.~K.} \bibnamefont{Miller}},
  \bibinfo{author}{\bibfnamefont{A.}~\bibnamefont{Cerezo}},
  \bibinfo{author}{\bibfnamefont{M.~G.} \bibnamefont{Hetherington}},
  \bibnamefont{and} \bibinfo{author}{\bibfnamefont{G.~D.~W.}
  \bibnamefont{Smith}}, \emph{\bibinfo{title}{Atom Probe Field Ion Microscopy}}
  (\bibinfo{publisher}{Clarendon Press}, \bibinfo{address}{Oxford},
  \bibinfo{year}{1996}).

\bibitem[{\citenamefont{Madey et~al.}(1999)\citenamefont{Madey, Nien, Pelhos,
  Kolodziej, Abdelrehim, and Tao}}]{Madey1999}
\bibinfo{author}{\bibfnamefont{T.~E.} \bibnamefont{Madey}},
  \bibinfo{author}{\bibfnamefont{C.-H.} \bibnamefont{Nien}},
  \bibinfo{author}{\bibfnamefont{K.}~\bibnamefont{Pelhos}},
  \bibinfo{author}{\bibfnamefont{J.~J.} \bibnamefont{Kolodziej}},
  \bibinfo{author}{\bibfnamefont{I.~M.} \bibnamefont{Abdelrehim}},
  \bibnamefont{and} \bibinfo{author}{\bibfnamefont{H.-S.} \bibnamefont{Tao}},
  \bibinfo{journal}{Surf. Sci.} \textbf{\bibinfo{volume}{438}},
  \bibinfo{pages}{191} (\bibinfo{year}{1999}).

\bibitem[{\citenamefont{Nien et~al.}(1999)\citenamefont{Nien, Madey, Tai,
  Leung, Che, and Chan}}]{Madey1999a}
\bibinfo{author}{\bibfnamefont{C.-H.} \bibnamefont{Nien}},
  \bibinfo{author}{\bibfnamefont{T.~E.} \bibnamefont{Madey}},
  \bibinfo{author}{\bibfnamefont{Y.~W.} \bibnamefont{Tai}},
  \bibinfo{author}{\bibfnamefont{T.~C.} \bibnamefont{Leung}},
  \bibinfo{author}{\bibfnamefont{J.~G.} \bibnamefont{Che}}, \bibnamefont{and}
  \bibinfo{author}{\bibfnamefont{C.~T.} \bibnamefont{Chan}},
  \bibinfo{journal}{Phys. Rev. B} \textbf{\bibinfo{volume}{59}},
  \bibinfo{pages}{10335} (\bibinfo{year}{1999}).

\bibitem[{\citenamefont{Fu et~al.}(2001)\citenamefont{Fu, Cheng, Nien, and
  Tsong}}]{FuChengNienTsong2001}
\bibinfo{author}{\bibfnamefont{T.-Y.} \bibnamefont{Fu}},
  \bibinfo{author}{\bibfnamefont{L.-C.} \bibnamefont{Cheng}},
  \bibinfo{author}{\bibfnamefont{C.-H.} \bibnamefont{Nien}}, \bibnamefont{and}
  \bibinfo{author}{\bibfnamefont{T.~T.} \bibnamefont{Tsong}},
  \bibinfo{journal}{Phys. Rev. B} \textbf{\bibinfo{volume}{64}},
  \bibinfo{pages}{113401} (\bibinfo{year}{2001}).

\bibitem[{\citenamefont{Lucier et~al.}(2005)\citenamefont{Lucier, Mortensen,
  Sun, and Gr\"{u}tter}}]{Lucier2005}
\bibinfo{author}{\bibfnamefont{A.-S.} \bibnamefont{Lucier}},
  \bibinfo{author}{\bibfnamefont{H.}~\bibnamefont{Mortensen}},
  \bibinfo{author}{\bibfnamefont{Y.}~\bibnamefont{Sun}}, \bibnamefont{and}
  \bibinfo{author}{\bibfnamefont{P.}~\bibnamefont{Gr\"{u}tter}},
  \bibinfo{journal}{Phys. Rev. B} \textbf{\bibinfo{volume}{72}},
  \bibinfo{pages}{235420} (\bibinfo{year}{2005}).

\bibitem[{\citenamefont{Rezeq et~al.}(2006)\citenamefont{Rezeq, Pitters, and
  Wolkow}}]{Rezeq2006}
\bibinfo{author}{\bibfnamefont{M.}~\bibnamefont{Rezeq}},
  \bibinfo{author}{\bibfnamefont{J.}~\bibnamefont{Pitters}}, \bibnamefont{and}
  \bibinfo{author}{\bibfnamefont{R.}~\bibnamefont{Wolkow}},
  \bibinfo{journal}{J. Chem. Phys.} \textbf{\bibinfo{volume}{124}},
  \bibinfo{pages}{204716} (\bibinfo{year}{2006}).

\bibitem[{\citenamefont{Bryl and Szczepkowicz}(2006)}]{BrylSzczepkowicz2006}
\bibinfo{author}{\bibfnamefont{R.}~\bibnamefont{Bryl}} \bibnamefont{and}
  \bibinfo{author}{\bibfnamefont{A.}~\bibnamefont{Szczepkowicz}},
  \bibinfo{journal}{Appl. Surf. Sci.} \textbf{\bibinfo{volume}{252}},
  \bibinfo{pages}{8526} (\bibinfo{year}{2006}).

\bibitem[{\citenamefont{Kuo et~al.}(2006)\citenamefont{Kuo, Hwang, Fu, Lin,
  Chang, and Tsong}}]{Kuo2006}
\bibinfo{author}{\bibfnamefont{H.-S.} \bibnamefont{Kuo}},
  \bibinfo{author}{\bibfnamefont{I.-S.} \bibnamefont{Hwang}},
  \bibinfo{author}{\bibfnamefont{T.-Y.} \bibnamefont{Fu}},
  \bibinfo{author}{\bibfnamefont{Y.-C.} \bibnamefont{Lin}},
  \bibinfo{author}{\bibfnamefont{C.-C.} \bibnamefont{Chang}}, \bibnamefont{and}
  \bibinfo{author}{\bibfnamefont{T.~T.} \bibnamefont{Tsong}},
  \bibinfo{journal}{Japanese Journal of Applied Physics}
  \textbf{\bibinfo{volume}{45}}, \bibinfo{pages}{8972} (\bibinfo{year}{2006}).

\bibitem[{\citenamefont{Fujita and Shimoyama}(2008)}]{FujitaShimoyama2008}
\bibinfo{author}{\bibfnamefont{S.}~\bibnamefont{Fujita}} \bibnamefont{and}
  \bibinfo{author}{\bibfnamefont{H.}~\bibnamefont{Shimoyama}},
  \bibinfo{journal}{Journal of Vacuum Science and Technology B}
  \textbf{\bibinfo{volume}{26}}, \bibinfo{pages}{738} (\bibinfo{year}{2008}).

\bibitem[{\citenamefont{Rahman et~al.}(2008)\citenamefont{Rahman, Onoda,
  Imaizumi, and Mizuno}}]{Rahman2008}
\bibinfo{author}{\bibfnamefont{F.}~\bibnamefont{Rahman}},
  \bibinfo{author}{\bibfnamefont{J.}~\bibnamefont{Onoda}},
  \bibinfo{author}{\bibfnamefont{K.}~\bibnamefont{Imaizumi}}, \bibnamefont{and}
  \bibinfo{author}{\bibfnamefont{S.}~\bibnamefont{Mizuno}},
  \bibinfo{journal}{Surf. Sci.} \textbf{\bibinfo{volume}{602}},
  \bibinfo{pages}{2128} (\bibinfo{year}{2008}).

\bibitem[{\citenamefont{Chang et~al.}(2009)\citenamefont{Chang, Kuo, Hwang, and
  Tsong}}]{Chang2009}
\bibinfo{author}{\bibfnamefont{C.-C.} \bibnamefont{Chang}},
  \bibinfo{author}{\bibfnamefont{H.-S.} \bibnamefont{Kuo}},
  \bibinfo{author}{\bibfnamefont{I.-S.} \bibnamefont{Hwang}}, \bibnamefont{and}
  \bibinfo{author}{\bibfnamefont{T.~T.} \bibnamefont{Tsong}},
  \bibinfo{journal}{Nanotechnology} \textbf{\bibinfo{volume}{20}},
  \bibinfo{pages}{115401} (\bibinfo{year}{2009}).

\bibitem[{\citenamefont{Kuo et~al.}(2008)\citenamefont{Kuo, Hwang, Fu, Lu, Lin,
  and Tsong}}]{Kuo2008}
\bibinfo{author}{\bibfnamefont{H.-S.} \bibnamefont{Kuo}},
  \bibinfo{author}{\bibfnamefont{I.-S.} \bibnamefont{Hwang}},
  \bibinfo{author}{\bibfnamefont{T.-Y.} \bibnamefont{Fu}},
  \bibinfo{author}{\bibfnamefont{Y.-H.} \bibnamefont{Lu}},
  \bibinfo{author}{\bibfnamefont{C.-Y.} \bibnamefont{Lin}}, \bibnamefont{and}
  \bibinfo{author}{\bibfnamefont{T.~T.} \bibnamefont{Tsong}},
  \bibinfo{journal}{Applied Physics Letters} \textbf{\bibinfo{volume}{92}},
  \bibinfo{pages}{063106} (\bibinfo{year}{2008}).

\bibitem[{\citenamefont{Hommelhoff et~al.}(2006)\citenamefont{Hommelhoff,
  Sortais, Aghajani-Talesh, and Kasevich}}]{Hommelhoff2006}
\bibinfo{author}{\bibfnamefont{P.}~\bibnamefont{Hommelhoff}},
  \bibinfo{author}{\bibfnamefont{Y.}~\bibnamefont{Sortais}},
  \bibinfo{author}{\bibfnamefont{A.}~\bibnamefont{Aghajani-Talesh}},
  \bibnamefont{and} \bibinfo{author}{\bibfnamefont{M.~A.}
  \bibnamefont{Kasevich}}, \bibinfo{journal}{Phys. Rev. Lett.}
  \textbf{\bibinfo{volume}{96}}, \bibinfo{pages}{077401}
  (\bibinfo{year}{2006}).

\bibitem[{\citenamefont{Hommelhoff et~al.}(2009)\citenamefont{Hommelhoff,
  Kealhofer, Aghajani-Talesh, Sortais, Foreman, and Kasevich}}]{Hommelhoff2009}
\bibinfo{author}{\bibfnamefont{P.}~\bibnamefont{Hommelhoff}},
  \bibinfo{author}{\bibfnamefont{C.}~\bibnamefont{Kealhofer}},
  \bibinfo{author}{\bibfnamefont{A.}~\bibnamefont{Aghajani-Talesh}},
  \bibinfo{author}{\bibfnamefont{Y.~R.~P.} \bibnamefont{Sortais}},
  \bibinfo{author}{\bibfnamefont{S.~M.} \bibnamefont{Foreman}},
  \bibnamefont{and} \bibinfo{author}{\bibfnamefont{M.~A.}
  \bibnamefont{Kasevich}}, \bibinfo{journal}{Ultramicroscopy}
  \textbf{\bibinfo{volume}{109}}, \bibinfo{pages}{423} (\bibinfo{year}{2009}).

\bibitem[{\citenamefont{Szczepkowicz and
  Ciszewski}(2002)}]{SzczepkowiczCiszewski2002}
\bibinfo{author}{\bibfnamefont{A.}~\bibnamefont{Szczepkowicz}}
  \bibnamefont{and}
  \bibinfo{author}{\bibfnamefont{A.}~\bibnamefont{Ciszewski}},
  \bibinfo{journal}{Surf. Sci.} \textbf{\bibinfo{volume}{515}},
  \bibinfo{pages}{441} (\bibinfo{year}{2002}).

\bibitem[{\citenamefont{Szczepkowicz and Bryl}(2004)}]{SzczepkowiczBryl2004}
\bibinfo{author}{\bibfnamefont{A.}~\bibnamefont{Szczepkowicz}}
  \bibnamefont{and} \bibinfo{author}{\bibfnamefont{R.}~\bibnamefont{Bryl}},
  \bibinfo{journal}{Surf. Sci. Lett.} \textbf{\bibinfo{volume}{559}},
  \bibinfo{pages}{L169} (\bibinfo{year}{2004}).

\bibitem[{\citenamefont{Szczepkowicz et~al.}(2005)\citenamefont{Szczepkowicz,
  Ciszewski, Bryl, Oleksy, Nien, Wu, and Madey}}]{Szczepkowiczetal2005}
\bibinfo{author}{\bibfnamefont{A.}~\bibnamefont{Szczepkowicz}},
  \bibinfo{author}{\bibfnamefont{A.}~\bibnamefont{Ciszewski}},
  \bibinfo{author}{\bibfnamefont{R.}~\bibnamefont{Bryl}},
  \bibinfo{author}{\bibfnamefont{C.}~\bibnamefont{Oleksy}},
  \bibinfo{author}{\bibfnamefont{C.-H.} \bibnamefont{Nien}},
  \bibinfo{author}{\bibfnamefont{Q.}~\bibnamefont{Wu}}, \bibnamefont{and}
  \bibinfo{author}{\bibfnamefont{T.~E.} \bibnamefont{Madey}},
  \bibinfo{journal}{Surf. Sci.} \textbf{\bibinfo{volume}{599}},
  \bibinfo{pages}{55} (\bibinfo{year}{2005}).

\bibitem[{\citenamefont{Szczepkowicz and Bryl}(2005)}]{SzczepkowiczBryl2005}
\bibinfo{author}{\bibfnamefont{A.}~\bibnamefont{Szczepkowicz}}
  \bibnamefont{and} \bibinfo{author}{\bibfnamefont{R.}~\bibnamefont{Bryl}},
  \bibinfo{journal}{Phys. Rev. B} \textbf{\bibinfo{volume}{71}},
  \bibinfo{pages}{113416} (\bibinfo{year}{2005}).

\bibitem[{\citenamefont{Antczak and Ehrlich}(2005)}]{AntczakEhrlich2005}
\bibinfo{author}{\bibfnamefont{G.}~\bibnamefont{Antczak}} \bibnamefont{and}
  \bibinfo{author}{\bibfnamefont{G.}~\bibnamefont{Ehrlich}},
  \bibinfo{journal}{Phys. Rev.} \textbf{\bibinfo{volume}{71}},
  \bibinfo{pages}{115422} (\bibinfo{year}{2005}).

\bibitem[{\citenamefont{Vurpillot et~al.}(2001)\citenamefont{Vurpillot, Bostel,
  and Blavette}}]{Vurpillot2001}
\bibinfo{author}{\bibfnamefont{F.}~\bibnamefont{Vurpillot}},
  \bibinfo{author}{\bibfnamefont{A.}~\bibnamefont{Bostel}}, \bibnamefont{and}
  \bibinfo{author}{\bibfnamefont{D.}~\bibnamefont{Blavette}},
  \bibinfo{journal}{Ultramicroscopy} \textbf{\bibinfo{volume}{89}},
  \bibinfo{pages}{137} (\bibinfo{year}{2001}).

\bibitem[{\citenamefont{Vurpillot et~al.}(2004)\citenamefont{Vurpillot, Cerezo,
  Blavette, and Larson}}]{Vurpillot2004}
\bibinfo{author}{\bibfnamefont{F.}~\bibnamefont{Vurpillot}},
  \bibinfo{author}{\bibfnamefont{A.}~\bibnamefont{Cerezo}},
  \bibinfo{author}{\bibfnamefont{D.}~\bibnamefont{Blavette}}, \bibnamefont{and}
  \bibinfo{author}{\bibfnamefont{D.}~\bibnamefont{Larson}},
  \bibinfo{journal}{Microsc. Microanal.} \textbf{\bibinfo{volume}{10}},
  \bibinfo{pages}{384} (\bibinfo{year}{2004}).

\bibitem[{\citenamefont{Niewieczerzal and Oleksy}(2006)}]{dn_czo}
\bibinfo{author}{\bibfnamefont{D.}~\bibnamefont{Niewieczerzal}}
  \bibnamefont{and} \bibinfo{author}{\bibfnamefont{C.}~\bibnamefont{Oleksy}},
  \bibinfo{journal}{Surf. Sci.} \textbf{\bibinfo{volume}{600}},
  \bibinfo{pages}{56} (\bibinfo{year}{2006}).

\bibitem[{\citenamefont{O.C.Zienkiewicz
  et~al.}(2005)\citenamefont{O.C.Zienkiewicz, R.L.Taylor, and
  J.Z.Zhu}}]{zienkiewicz}
\bibinfo{author}{\bibnamefont{O.C.Zienkiewicz}},
  \bibinfo{author}{\bibnamefont{R.L.Taylor}}, \bibnamefont{and}
  \bibinfo{author}{\bibnamefont{J.Z.Zhu}}, \emph{\bibinfo{title}{Finite Element
  Method: Its Basis and Fundamentals}} (\bibinfo{publisher}{Butterworth
  Heinemann}, \bibinfo{address}{Burlington}, \bibinfo{year}{2005}).

\bibitem[{\citenamefont{Rahman et~al.}(2006)\citenamefont{Rahman, Gorman,
  Barnes, Williams, and Langtangen}}]{Rahman2006}
\bibinfo{author}{\bibfnamefont{S.}~\bibnamefont{Rahman}},
  \bibinfo{author}{\bibfnamefont{J.}~\bibnamefont{Gorman}},
  \bibinfo{author}{\bibfnamefont{C.~H.~W.} \bibnamefont{Barnes}},
  \bibinfo{author}{\bibfnamefont{D.~A.} \bibnamefont{Williams}},
  \bibnamefont{and} \bibinfo{author}{\bibfnamefont{H.~P.}
  \bibnamefont{Langtangen}}, \bibinfo{journal}{Phys. Rev. B}
  \textbf{\bibinfo{volume}{73}}, \bibinfo{pages}{233307}
  (\bibinfo{year}{2006}).

\bibitem[{get()}]{getfem}
\emph{\bibinfo{title}{Getfem++ home page: http://home.gna.org/getfem}}.

\bibitem[{tet()}]{tetgen}
\emph{\bibinfo{title}{Tetgen home page: http://tetgen.berlios.de/}}.

\bibitem[{\citenamefont{Frenkel and Smith}(1996)}]{frenkel_MD}
\bibinfo{author}{\bibfnamefont{D.}~\bibnamefont{Frenkel}} \bibnamefont{and}
  \bibinfo{author}{\bibfnamefont{B.}~\bibnamefont{Smith}},
  \emph{\bibinfo{title}{Understanding Molecular Simulation}}
  (\bibinfo{publisher}{Academic Press}, \bibinfo{address}{San Diego},
  \bibinfo{year}{1996}).

\bibitem[{\citenamefont{Nien and Madey}(1997)}]{NienMadey1997}
\bibinfo{author}{\bibfnamefont{C.-H.} \bibnamefont{Nien}} \bibnamefont{and}
  \bibinfo{author}{\bibfnamefont{T.~E.} \bibnamefont{Madey}},
  \bibinfo{journal}{Surf. Sci. Lett.} \textbf{\bibinfo{volume}{380}},
  \bibinfo{pages}{L527} (\bibinfo{year}{1997}).

\end{thebibliography}

\end{document}